\documentclass[12pt]{article}
\usepackage{amssymb}

\makeatletter
\renewcommand\section{\@startsection {section}{1}{\z@}%
                                 {-3.5ex \@plus -1ex \@minus -.2ex}
                                   {2.3ex \@plus.2ex}%
                                   {\normalfont\large\bfseries}}
\renewcommand\subsection{\@startsection{subsection}{2}{\z@}%
                                   {-3.25ex\@plus -1ex \@minus -.2ex}%
                                     {1.5ex \@plus .2ex}%
                                     {\normalfont\bfseries}}
\renewcommand\subsubsection{\@startsection{subsubsection}{3}{\z@}%
                                   {-3.25ex\@plus -1ex \@minus -.2ex}%
                                     {1.5ex \@plus .2ex}%
                                     {\normalfont\itshape}}
\makeatother

\newcounter{multieqs}

\newcommand{\cA}{{\cal A}}

\newcommand{\cG}{{\cal G}}

 \newcommand{\cN}{{\cal N}}

 \newcommand{\cZ}{{\cal Z}}
 \newcommand{\bo}{{\bar o}}

\newcommand{\bz}{{\bar z}}

\newcommand{\bS}{{\bar S}}

\def\e{\epsilon}    \def\i{{\rm
i}}   \def\l{\lambda} 
    
\def\s{\sigma}    
  \def\G{\Gamma}

 \def\ZZ{\mZ}
\def\bo{{\raise.15ex\hbox{\large$\Box$}}}  
\def\face{{\raise.2ex\hbox{$\displaystyle \bigodot$}\mskip-2.2mu \llap
{$\ddot \smile$}}}             

\def\leftrightarrowfill{$\mathsurround=0pt \mathord\leftarrow
        \mkern-6mu \cleaders\hbox{$\mkern-2mu \mathord-
        \mkern-2mu$}\hfill
        \mkern-6mu \mathord\rightarrow$}       
\def\dvec#1{\vbox{\ialign{##\crcr
        \leftrightarrowfill\crcr\noalign{\kern-1pt\nointerlineskip}
        $\hfil\displaystyle{#1}\hfil$\crcr}}}     

\def\beq{\begin{equation}} \def\eeq{\end{equation}}

\def\beqx{\begin{displaymath}} \def\eeqx{\end{displaymath}}

\def\beql{\begin{eqnarray}} \def\eeql{\end{eqnarray}}

\newcommand{\bea}{\begin{eqnarray}} \newcommand{\eea}{\end{eqnarray}}

 \newcommand{\R}[1]{(\ref{eq:#1})}
\newcommand{\mod}{\;\;\;\;{\rm mod }\;}

\def\non{\nonumber \\}  \def\[{\left [} \def\]{\right ]}
\def\({\left (} \def\){\right )} \def\C{\Gamma}

\def\ZZ{\mathbb{Z}} \def\RR{\mathbb{R}} 

\def\eps{\epsilon}  
   
\def\lam{\lambda}  \def\sig{\sigma}
  
\def\ome{\omega}

\def\cA{{\cal A}}      \def\cG{{\cal G}}
      \def\cN{{\cal N}}
      
    \def\cZ{{\cal Z}}

 \def\ove#1{\frac{1}{#1}}

 \def\+{\oplus}

\begin{document}

\thispagestyle{empty}
\begin{flushright}
\parbox[t]{2in}{CU-TP-1087\\KUL-TF-2003/23\\
  hep-th/0306091}
\end{flushright}

\vspace*{0.4in}

\begin{center}
{\Large \bf 
BPS orientifold planes from crosscap states in Calabi-Yau compactifications
} 

\vspace*{0.3in}
L.~Huiszoon\footnote{Email address: LennaertRen.Huiszoon@fys.kuleuven.ac.be}~
and~K.~Schalm\footnote{Email address: kschalm@physics.columbia.edu}
\\[.3in]

${}^1${\em Instituut voor Theoretische Fysica\\
Katholieke Universiteit Leuven\\
 Celestijnenlaan 200D\\
B-3001 Leuven}\\[0.1in]
${}^{2}${\em Department of Physics \\
Columbia University \\
New York, NY 10027}\\[.2in]

{\bf Abstract}
\end{center}

\noindent
We use the results of hep-th/0007174 on the simple current classification
of open unoriented CFTs
 to construct half supersymmetry
  preserving crosscap states for rational Calabi-Yau compactifications.
We show that the corresponding orientifold fixed planes obey
the BPS-like relation $M= e^{\i\phi} Q$. To prove this relation, it is
  essential that the worldsheet 
CFT properly
includes the degrees of freedom from the uncompactified space-time
component. The BPS-phase $\phi$ can be identified with 
the 
  automorphism type of 
the crosscap states.
To illustrate the method we compute crosscap states in 
Gepner models with each $k_i$ odd.

\newpage
\setcounter{footnote}{0}

\section{Introduction}

In this letter we present a universal class of crosscap states for
$\cN=2$ rational CFTs corresponding to strings on Calabi-Yau
manifolds. A generic property of a 'rational' Calabi-Yau
compactification is that 
the worldsheet theory needs to be GSO projected to ensure space-time
supersymmetry. Such a GSO projection is
equivalent to a 
so called simple current extension by the spectral flow current $S$ of
the underlying 
chiral symmetry algebra of the worldsheet CFT.
In~\cite{FOE} consistent boundary and crosscap states for
arbitrary simple current 
modular invariants, of which simple current
extensions are a subset, have been constructed.
We can therefore apply the results of~\cite{FOE}
to the specific 
case of a CY compactification.
The connection
between the simple 
current extension with $S$ and space-time supersymmetry is reflected in
that these simple current boundary and crosscap states preserve half
the space-time supersymmetry. Half supersymmetric boundary states in CY
compactifications,
i.e. BPS D-branes, have been extensively studied in the past few years,
both from the orbifold (see
e.g. \cite{Recknagel:1997sb,Brunner:1999jq,Douglas:2001ug} and 
citations thereof) and simple current point of
view (see e.g. \cite{Fuchs:2000gv,Fuchs:2000fd,walcher}). Consistent
type I 
 CY compactifications with D-branes, however, 
need half-supersymmetry preserving
 orientifold planes. It is the corresponding crosscap states
 we discuss here. 

In section 2 we will give an intuitive explanation of 
the results of~\cite{FOE} for $\ZZ_N$ simple
current invariants. We will for simplicity assume that the currents 
do not have fixed points. In the presence of fixed points, the
formulas for the boundary 
states change qualitatively \cite{Fuchs:2000gv, Fuchs:2000fd}, 
whereas those for the
crosscap states do not. Our focus is here on the
latter and we believe that the inclusion
 of fixed points will not change our main results. 
In section 3 we then apply these
formulas to `rational' CY compactifications. 
We show that, when the uncompactified part of space-time
is properly taken into account, the O-planes are BPS-like, i.e. they
obey 
a mass-charge  
relation $M = e^{\i\phi} Q$, and we determine the phase $\phi$ 
in terms of CFT
quantities.  The BPS-like relation 
is a consequence of preserving half of the space-time supersymmetries. 
Interestingly, for D-branes the 
BPS property can be derived from the boundary
state without reference 
to the uncompactified part of the theory~\cite{walcher}. 
We explain at the
mathematical level why this is so. Finally we briefly illustrate these
methods with the computation of the mass and charges of crosscap
states in Gepner models with each $k_i$ odd.

Despite the required presence of crosscap states in consistent type I
CY 
compactifications, the interest in unoriented $\cN=2$ CFTs 
has begun rather
recently~\cite{Acharya:2002ag,Blumenhagen:2002wn,Brunner:2003zm,Misra:2003zv,Diaconescu:2003dq}.  
Using  
a more geometrical approach based on linear sigma
models,~\cite{Brunner:2003zm} found the locations of CY-orientifold
planes as
fixed points of anti-holomorphic
\cite{Blumenhagen:2002wn,Misra:2003zv} or holomorphic isometries. To
understand type I CY compactifications, we wish to
know their physical characteristics, such as charge and tension, as
well. At rational points in the CY moduli space, the method described
here can be used to determine these.

In the midst of this project we noticed the posting of
the conference proceedings 
\cite{Govindarajan:2003vv} to the archive, in which BPS crosscap states 
for Gepner Models are constructed by exploiting the phase symmetries
(see also the subsequent article \cite{Govindarajan:2003vp}).
Since phase symmetries are realized by simple currents, the results of
\cite{Govindarajan:2003vv} are in agreement with our results. The
power of the 
construction proposed here is that it is applicable to any rational
$N=2$ SCFT that describes a CY compactification.

\section{Boundary and crosscap states for simple current 
    extensions} 
\label{sec:bound-crossc-stat}

Consider a CFT with the same left- and right-chiral algebra $\cA$ and
a C-diagonal torus partition function (i.e. the modular invariant
theory which pairs left-movers with right-moving Charge Conjugates).
A complete set of
boundary states that preserve a diagonal subalgebra of the left-right
symmetry algebra  
$\cA \times \bar{\cA}$ is given by the Cardy states~\cite{cardy}
\beq
\label{eq:cardy}
|B_a \rangle = \sum_i \frac{S_{ia}}{\sqrt{S_{i0}}} |i\rangle\rangle_1~,
\eeq
where $S_{ia}$ is the modular $S$-matrix of $\cA$ and 
the sum is over all primaries.
The Ishibashi state
$|i\rangle\rangle_1$ is a 
coherent state of all C-diagonal closed string states in sector
$i$. The boundary state  
preserves a diagonal subalgebra, which means that all states
contribute with the same weight, namely $1$, to $|i\rangle\rangle_1$. 
The boundary label $a$ runs over all primaries of $\cA$. Different
labels can be thought
of as labeling branes wrapping different cycles. 

Fuchs and Schweigert extended this result of Cardy and constructed
boundary states for theories whose
modular invariant is a simple current extension~\cite{SBB}.
Recall~\cite{simple} that
simple currents $J$ are primary fields whose fusion with any other
primary $i$
field yields a {\em single} field $j=Ji$. 
Integer conformal weight 
simple currents can be used to extend the chiral algebra 
$\cA \stackrel{J}{\subset} \cA^{ext}$. Under this extension primary
fields arrange themselves into orbits $[i] =
\left\{i,Ji,J^2i,\ldots\right\}$. Orbits with integer monodromy charge
$Q_J(i)\equiv h_i+h_J-h_{Ji}$ mod 
$\ZZ$ under $J$, are the primaries of $\cA^{ext}$. Non-integer charged
fields are 
projected out. Such an extension 
is often referred to as a `simple current orbifold'.
Although not quite correct from the worldsheet point of view,\footnote{The
  difference emphasized in the introduction is that a
  worldsheet-orbifold makes a chiral algebra smaller, whereas an 
extension makes it larger. Orbifolds and simple current extensions are
in fact each others
inverse.} this terminology sometimes makes sense from the point of
view of the target space.  
 In WZW models based on Lie
group $G$ for example, an extension
by a simple current group $\ZZ_N$ amounts geometrically to strings
moving on 
$G/\ZZ_N$.

With this geometric picture in mind, a natural guess for 
the boundary state of a $\ZZ_N$ extension is a sum over 'images',
\beq
\label{eq:BBB}
|B_{[a]}\rangle = \ove{\sqrt{N}}\sum_{n=0}^{N-1} |B_{J^na}\rangle~,
\eeq
where $J$ is a simple current that generates the $\ZZ_N$. (The
normalization follows from CFT arguments, see
eq.~\R{S} below.)
An important simple current identity for the modular
$S$-matrix~\cite{simple, intril},
\beq \label{eq:SJ}
S_{Ji,j} = e^{2\pi\i Q_{J}(j)} S_{i,j}~,
\eeq
allows the boundary state to be written as
\beq \label{eq:BS}
|B_{[a]}\rangle = \sqrt{N} 
\sum_{\{i|REP_{[i]},~Q_J(i)=0\}}
 \frac{S_{ia}}{\sqrt{S_{i0}}} \; |[i]\rangle\rangle_{Q_J(a)}~.
\eeq
Here the sum is over representatives of chargeless $J$-orbits and
\beq
|[i]\rangle\rangle_{Q_J(a)} = \sum_{n=0}^{N-1} e^{2\pi\i n Q_{J}(a)}|J^ni\rangle\rangle_1~.
\eeq
This natural guess indeed corresponds to the boundary states
constructed by Fuchs and Schweigert. Moreover, these boundary states
are 'more general' than the Cardy state~\R{cardy}
in the following sense.
Recall that the primaries $(i,Ji,J^2i,...,J^{N-1}i)$ with $Q_J(i)=0$ group into one
primary $[i]$ of the extended algebra $\cA^{ext}$. We see that 
on the boundary state $|B_{[a]}\rangle$ the closed strings in sector $i$ and
$Ji$ are reflected with a relative phase $e^{2\pi\i Q_{J}(a)}$. From the point of view of
the extended algebra $\cA^{ext}\times \bar{\cA}^{ext}$, 
the boundary state
therefore does not respect a diagonal subalgebra, but a twisted one. The boundaries are
said to obey the {\em twisted gluing condition}
\beq
[J_n - (-1)^{h_J} e^{-2\pi\i Q_{J}(a)} \bar{J}_{-n}] |B_{[a]}\rangle = 0 \;\;\;  ,
\eeq 
with respect to the simple current $J \in \cA^{ext}$.
The phase $e^{-2\pi\i Q_{J}(a)}$ is the {\em automorphism type} of the
boundary. Note that for a $\ZZ_N$ simple current $J$, the monodromy
charge is a fraction of $N$:
\beq
 Q_J(a) = \frac{n}{N}~,~~~~n \in \ZZ ~.
\eeq
Thus
the automorphism type takes values in $\ZZ_N$. In particular 
for $Q_J(a)=0$ mod $1$ the
boundary state~\R{BS} preserves the diagonal subalgebra. By
construction this is the usual Cardy state for $\cA^{ext}$. We
infer therefore --- correctly --- 
that the modular S-matrix of the extended theory is
expressible in terms of the modular S-matrix of the original theory
\beq
  \label{eq:S}
  S_{[a][b]} =  N S_{ab} \;\;\; .
\eeq
The $N$-dependence in this relation 
explains the  normalization choice in eq.~\R{BBB}.

Pradisi, Sagnotti and Stanev~\cite{PSS}
(PSS) found the formula
analogous to Cardy's for crosscap states:~\footnote{See~\cite{Angelantonj:2002ct} for a review.}
\beq
|\C \rangle^{\s(0)}_0 = \s(0)\sum_i \frac{P_{i0}}{\sqrt{S_{i0}}}
|i\rangle\rangle_{1,C}~. 
\eeq
Here $P$ is the pseudo-modular matrix $P=\sqrt{T}ST^2S\sqrt{T}$ built
from the modular $T$- and $S$-matrices,   
the sum runs over all primaries and $0$ denotes the vacuum
representation. In addition $\s(0)$ is 
the undetermined sign in the M\"obius
strip,  which is ultimately fixed by tadpole cancellation. The
crosscap Ishibashi state is similar to boundary Ishibashi states,
except that even/odd 
levels contribute with opposite signs. A modified crosscap exists for
every simple current 
$K$ of $\cA$, given by~\cite{klein}\footnote{Note that the crosscap
  coefficient 
in~\cite{klein} is written slightly different, namely
$\G'_i=\frac{P_{iK}}{\sqrt{S_{iK}}}$.  
In~\cite{NunoBert} it is argued that
$\G_i=\frac{P_{iK}}{\sqrt{S_{i0}}}$
is the correct crosscap state and that the Klein bottle
 is calculated as $K_m= \sum_i \G_i\G_i S_{im} e^{2\pi\i Q_K(i)}$,
 where $e^{2\pi\i 
 Q_K(i)}$ is due to the action of $\Omega$ on the Ishibashi
states. Similar remarks apply 
 to the boundary state.}
\beq \label{eq:KBC}
|\C \rangle^{\s(K)}_K = \s(K) \sum_i \frac{P_{iK}}{\sqrt{S_{i0}}}
|i\rangle\rangle_{1,C}~. 
\eeq
The primary $K$ is called the Klein bottle current (KBC). In WZW
models~\cite{HSS} it was 
shown that the label $K$ plays a role
that is similar to the boundary label, namely different labels $K$
represent O-planes at different locations. A natural guess for the
 crosscap state in simple current extensions is therefore
~\cite{FOE,T,thesis}
\beq \label{eq:CROSS}
|\C \rangle^{[\s]}_{[K]} = \ove{\sqrt{N}}
\sum_{n=0}^{N-1}|\C \rangle^{\s(J^nK)}_{J^nK}~.
\eeq
As we will see the signs $\s(J^nK)$ are not completely arbitrary. 

In the case of boundary states, the important identity (\ref{eq:SJ})
allowed us to 'perform' the sum over images. 
For the $P$-matrix, however, the analogous 
 simple current identity is at first sight very
different~\cite{thesis}
\beq \label{eq:PJ}
 P_{J^2i,j} = \e_{J^2}(i) e^{2\pi\i [Q_{J}(j)-Q_J(Ji)]} P_{i,j}~, 
\eeq 
where 
\beq
\e_J(i) := e^{\pi\i [h_i - h_{Ji}]}~.
\eeq
Most importantly, for $N$ even, this relation is truly {\em qualitatively} different
than its analogue~\R{SJ}. In that case only $P$-matrix elements on the same $J$-orbit 
that differ by {\em two} steps are related. From now on we focus on the interesting case
$N$ even. (This is also the case relevant for CY compactifications.)

Suppose first that the signs $\s(J^nK)$ in~\R{CROSS} are such
that\footnote{Note that for an integer spin current
$\eps^2_{J^{2m}}(K)e^{-4\pi\i Q_{J^m}(J^mK)} =
e^{2\pi\i[Q_{J^{2m}}(K)-2Q_{J^m}(J^mK)]} = 1$, therefore
$\sig_0,~\sig_1$ are indeed signs.}
\bea \label{eq:signs}
\s(J^{2m}K) \e_{J^{2m}}(K) e^{-2\pi\i Q_{J^m}(J^mK)} & = & \s_0 \;\;\;,
\non
\s(J^{2m+1}K) \e_{J^{2m}}(JK) e^{-2\pi\i Q_{J^m}(J^mJK)} & = & \s_1
\;\;\;,  
\eea
for any pair of signs $\s_0,\s_1$. As we show in appendix A, these
choices ensure that only $Q_J(i)=0$ primaries, 
i.e. fields of $\cA^{ext}$, 
couple to the crosscap. With this choice
the crosscap state equals
\beq \label{eq:Cross}
|\C \rangle^\s_{[K]} = \sqrt{N} \sum_{\{i|REP_{[i]},~Q_J(i)=0\}} 
\(\frac{\s_0 P_{iK} + \s_1 P_{i,KJ}}{2 \sqrt{S_{i0}}}\)
 |[i]\rangle\rangle_{\e_J(K)\s}~,
\eeq
where $\s=\s_0/\s_1$ and 
\beq \label{eq:twist}
|[i]\rangle\rangle_{\e_J(K)\s} = \sum_{n=0}^{N-1} [\e_J(K)\s]^n \e_{J^n}(i) 
|J^ni\rangle\rangle_{1,C}~.
\eeq
Therefore, the crosscap states obey a twisted gluing
condition,
\beq \label{eq:glue}
[J_n - (-1)^{n + h_J} \e^{\ast}_J(K)\s \bar{J}_{-n}] |\C \rangle^\s_{[K]} = 0~.
\eeq 
and $\e^{\ast}_J(K)\s = (\e_J(K)\s)^{-1}$ 
is the automorphism type of the crosscap state. 
Note that the sign $\e_{J^n}(i) = e^{\pi\i[h_i - h_{J^ni}]}$ in~\R{twist}
is the parity of the level of the descendant $J^ni$ in the module
$[i]$.    
To make sure that the overall sign of the crosscap Ishibashi state
$|[i]\rangle\rangle_{\e_J(K)\s}$ is correct, we must insist that the
representative $i$ has lowest conformal weight (mod $2$) in the orbit
$(i,Ji,...,J^{N-1}i)$. From now on we assume that this is the case.

Not any Klein bottle current is allowed. The current $K$ must have
monodromy charge 
\beq
\label{eq:even}
Q_J(K) = \frac{2p}{N}~,~~p\in \ZZ~,
\eeq
otherwise~\R{signs} cannot be solved over signs 
\cite{thesis}. As a consequence, $[\e_J(K)\s]^N=1$ and the
automorphism type of the planes takes values in $\ZZ_N$. The
automorphism types of simple current D-branes and orientifold planes 
are
therefore the {\em same}, as we indeed geometrically expect, even though the
responsible CFT mechanisms are different.

Suppose we would make sign choices in~\R{CROSS} other
than~\R{signs}. In that case only charged fields would couple to the 
crosscap.
Charged fields are not representations of the simple current extended
algebra, and there is no linear combination of $J$ and $\bar{J}$ that is preserved by the
crosscap.
 Such
crosscap states are not obviously inconsistent, but beyond the scope of this letter and we will proceed with~\R{signs} and
the result~\R{Cross}.

Again, when the automorphism type is trivial, i.e. $\e_J(K)\s = 1$, we
expect~\R{Cross} to be a PSS+KBC crosscap state for the extended
algebra $\cA^{ext}$ by construction. We infer that the $P$-matrix of
a simple current extension is related to the original $P$-matrix as~\cite{T,thesis}
\beq                                       \label{eq:Pext}
P_{[i][j]} = \sum_{n=0}^{N-1} \e_{J^n}(i) P_{J^{n}i,j} = \frac{N}{2} [P_{ij} +
\e_{J}(i)P_{Ji,j}] ~,
\eeq
where we used eq. (\ref{eq:PJ}) in the last step. 
Indeed, this expression is correct when the
representatives $i$ and $j$ are chosen to have the lowest conformal weight
(mod $2$) in their orbits (and the currents have no fixed points.) For later
purposes we also need the $P$-matrix for non-cyclic simple current groups
$\cG$. The generalization is  straightforward:
\beq \label{eq:PextG}
P_{[i][j]} = \sum_{J\in \cG} \e_{J}(i) P_{Ji,j}~.
\eeq

To conclude: we have shown that boundary and crosscap states of $\ZZ_N$ simple current
invariants of $\cA$ posses a $\ZZ_N$-automorphism type with respect to the current that
extends $\cA$ to $\cA^{ext}$. When $N$ is even the boundary and crosscap states of 
the extension have a qualitatively different feature: the boundary coefficient of boundary
$[a]$ is the boundary coefficient of a representative $a$. In contrast to this, the
crosscap coefficient of $[K]$ is the sum of the crosscap coefficient with Klein bottle
currents $K$ and $JK$.

\section{Crosscap states for Calabi-Yau compactifications} \label{sec-CY}
\setcounter{equation}{0}

In the previous section we have reviewed the theory of boundary and crosscap states in
simple current extension invariants. 
In string theory, simple currents play a
role in at least three steps in the construction of realistic models: (i) simple currents 
implement field identification and selection rules in coset constructions, (ii)
 simple currents realize alignment of spin structures in tensor product CFTs and (iii)
 they implement GSO projections. We will apply 
 the general theory outlined above to the last step in the construction, the GSO
 projection.\footnote{The reason we only apply the theory to the last
   step is that $\cN=2$ worldsheet susy does not allow boundary or
   crosscap states with non-trivial automorphism type in alignment or
   identification currents. Hence the Cardy-PSS solution is sufficient for 
the first two extensions~\cite{Fuchs:2000gv}.}

\subsection{The bulk theory}

We start by reviewing the closed string sector of Calabi-Yau compactifications in the
language of CFT. We should stress that this construction is only appropriate at rational 
points of the CY moduli space. These are `very symmetric' points in the space of
CY deformations where the $2d$ symmetry algebra is enlarged to a theory with only a finite
number of primary fields (with respect to the Virasoro plus extended algebra). Examples
are the Fermat polynomial representations 
of Calabi-Yau manifolds as vanishing loci in
weighted projected space. 

The chiral algebra at a rational point of a type II compactification
on a CY $n$-fold can be constructed as follows. The starting point in
the construction of the worldsheet theory is a tensor product 
\beq
\label{eq:ten} 
D_{8-n,1} \otimes \cA_{3n} \;\;\; 
\eeq 
where $D_{r,1}$
is the affine algebra based on $SO(2r)$ at level one and $\cA_{3n}$
any rational chiral bosonic subalgebra of a $\cN=2$ superconformal
algebra.  The latter contains a Virasoro algebra with conformal
anomaly $3n$ and a $U(1)_{R}$ algebra and has two distinguished
simple currents.  The {\em supercurrent} $v$ has order $2$ and spin
$h_v=3/2$.  States with integer (half-integer) monodromy charge with
respect to the supercurrent are in the NS (R) sector. The second
simple current is called {\em spectral flow} $s$. It has spin
$h_s=c/24 = n/8$ and model dependent order $N_s$. The
monodromy charge $Q_s(\l) = q_\l/2$ mod $1$ equals half the $U(1)_R$ charge $q_\l$ of $\l$.
 
The $D_{8-n,1}$ factor in~\R{ten} describes the uncompactified part of
the theory.  The use of the unitary group 
$SO(16-2n)$ rather the non-unitary group
$SO(10-2n,2)$, corresponding to the Lorentz group of the space-time fermions plus
bosonized superghosts, is called {\em the bosonic string map}
\cite{BSM} (for a review see~\cite{LSW}).
 The $D_{8-n,1}$ theory has four primaries, $X_{8-n} \in
(O_{8-n},S_{8-n},V_{8-n},C_{8-n})$ with conformal weights
$(0,(8-n)/8,1/2,(8-n)/8)$ that realize a simple current group $\ZZ_2
\times \ZZ_2$ when $n$ is even and $\ZZ_4$ when $n$ is odd. The vector
$V_{8-n}$ plays the role of the supercurrent.
 The singlet and vector are in the NS sector and the spinor and
conjugate spinor are in the R sector. In order to read off
the string spectrum from the partition functions, one has to
perform the (inverse) bosonic string map~\cite{BSM}: 
\beq
\{O_{8-n},S_{8-n},V_{8-n},C_{8-n}\} \rightarrow
\{V_{4-n},-S_{4-n},O_{4-n},-C_{4-n}\} \;\;\; .  
\eeq
Note that the current
$(S_{8-n},s)$ has conformal weight $1$ and can therefore be used to extend the
chiral algebra. This is one of the reasons for using $SO(16-2n)$
instead of the expected little group $SO(8-2n) \subset SO(10-2n,2)$.

The
algebra~\R{ten} cannot describe a $\cN=2$ superconformal theory since
the spin structures (R or NS) of the space-time and internal part are
not 
aligned.  A
superconformal theory is obtained when we extend~\R{ten} by
$(V_{8-n},v)$. Let us denote this 
extended algebra by $\cA^{ws}$. The
primaries of $\cA^{ws}$ are $[X_{8-n},\l]$ and
are subject to the following identification and selection rules: 
\beq
[X_{8-n},\l] \sim [V_{8-n}X_{8-n},v\l] \;\;\; , \;\;\; Q_v (\l) =
Q_{V_{8-n}} (X_{8-n}) \mod 1\;\;\; .  
\eeq 
By the identification, the
order $N_S$ of the spectral flow current $S\equiv
[S_{8-n},s]$ is either $N_s$ or $N_s/2$, depending on the model under
consideration. The supercurrent of $\cA^{ws}$ is $V\equiv [O_{8-n},v]$
and the vacuum is $O \equiv [O_{8-n},0]$. Note however that $N_S \in 2\ZZ$ because
$S_{8-n}^{2}=O_{8-n}$ for $n$ even and $S_{8-n}^{4}=O_{8-n}$ for $n$
odd.

We can use the spectral flow current to build a simple current modular
invariant $\cZ(\cA^{ws},S)$. This procedure is analogous to the GSO
projection and the resulting CFT describes closed oriented strings on
$\RR^{1,9-n}\otimes CY_n$.  This CFT has fields $S(z)$ and $\bS(\bz)$
whose zero modes $S_0$ and $\bar{S}_0$ generate space-time supersymmetry
transformations. Because $S$ has integer spin, 
the invariant
$\cZ(\cA^{ws},S)$ can be thought of as a C-diagonal 
invariant of a larger algebra $\cA^{ext }$ obtained from $\cA^{ws}$ by
an extension by $S$.  This section can therefore be summarized by the
following sequence of embeddings 
\beq 
\label{eq:embed}
D_{8-n,1} \otimes \cA_{3n}
\stackrel{{(V_{8-n},v)}}{\subset}\cA^{ws}
\stackrel{{S}}{\subset} \cA^{ext } \;\;\; .  
\eeq 
where we
have indicated which simple currents are used as extensions. 
Note that this split-up is different from the one used
in~\cite{Fuchs:2000gv}. The guiding philosophy there was that the
properties of branes on the CY manifold, i.e. boundary states in 
the internal CFT should be quite independent of
the external space-time. This led the authors to 
first extend $\cA_{3n}$ ($\cA^{wsusy}$ in their notation)
by the current $u\equiv v^ns^2$ to $\cA^{cy}$; a half-way
  GSO-projection onto integer $U(1)_R$ charges (integer worldsheet
  fermion number). The latter theory $\cA^{cy}$ contains the fields 
$o_{cy},~s_{cy},~v_{cy}$ and $c_{cy}$ --- all simple currents --- 
with the same fusion rules as
the primary fields of $D_{8-n,1}$. A second extension by these
currents yields $\cA^{ext }$. Because the latter extension acts without fixed points, 
the boundary states of $\cA^{cy}$ and $\cA^{ext }$ indeed have the same qualitative
features. 
However this latter extension is a $\ZZ_{even}$ extension and therefore the crosscap
states of $\cA^{cy}$ and $\cA^{ext }$ are qualitatively different, as explained in the
previous section. In particular, the crosscap state of $\cA^{ext }$ is the sum of the PSS
crosscap state and the PSS+KBC crosscap state of $\cA^{cy}$, where the Klein bottle
 current is $S$. We will see this in more detail in the next subsection.
 
\subsection{A BPS condition for O-planes}

Using the general theory reviewed in section 2, 
it is now straightforward to write down
 a consistent set of crosscap states for rational CY
compactifications. The theory $\cA^{ext}$, describing strings on CY
and obtained from $\cA^{ws}$ by extension with $S$, has crosscap
states   
\beq \label{eq:Cr}
|\C \rangle^\s_{[K]} = \sqrt{N_S} \sum_{\{i|REP_{[i]},~Q_S(i)=0\}} \(\frac{\s_0 P_{iK} + \s_1
P_{i,KS}}{2\sqrt{S_{0i}}}\)
 |[i]\rangle\rangle_{\e_S(K)\s} \;\;\; .
\eeq
where $K$ a simple current of $\cA^{ws}$ and $\s=\s_0/\s_1$. For $K=O$ we see indeed that this crosscap state is
the sum of a PSS crosscap state and the PSS+KBC crosscap state, where the Klein bottle
 current is $S$. 
 This crosscap state can therefore have non-trivial automorphism
  type with respect to the spectral flow current $S$. With respect to 
all other
  currents of $\cA^{ext}\times \bar{\cA}^{ext}$
it preserves the {\em
    diagonal} subgroup. One of these currents is the $U(1)_R$
  current. The corresponding orientifold-planes are therefore all of
  A-type \cite{Ooguri:1996ck,Brunner:2003zm}.
 
 From this formula we can derive some basic properties of the corresponding orientifold
 fixed planes. One should keep in mind that a single crosscap state, specified by
 the triple $(K,\s_0,\s_1)$ represents in general a configuration of
 O-planes (WZW models are an example \cite{HSS}).   
 \begin{itemize}
\item
  The gluing condition~\R{glue} implies that these planes preserve
 one half of space-time supersymmetry generated by the linear
  combination $S_0 + \e^{\ast}_S(K)\s
  \bar{S}_0$. 

 \item The O-planes are BPS-like.
 From the expansion~\R{Cr} one can read off the
 charges of the O-planes with respect to closed strings $|j;\bar{j}\rangle$ 
 by calculating the
 overlap $\langle j;\bar{j} |\C \rangle^\s_{[K]}$. 
  The mass $M$ and central charge $Q$ of the planes are given by
 \beq
M = \langle O;\bar{O}  |\C \rangle^\s_{[K]} \;\;\; , \;\;\; Q = \langle S;\bar{S}  |\C \rangle^\s_{[K]}
 \eeq  
 where $\langle O;\bar{O} |$ denotes the graviton and $\langle S;\bar{S}|$
  is the top RR form \cite{walcher}. 
This is the chiral-chiral outstate
$\langle S;\bar{S}|$ with both left- and right $U(1)_R$ charge
 maximal, $q_{U(1)_R} = c/6$, and is
 obtained by acting 
on the vacuum with precisely the simple currents $S$ and $\bar{S}$.
From the formula for the twisted Ishibashi states~\R{twist} we then
 immediately  infer that the
  planes obey a BPS-like relation 
\beq
M = \s \e_S(K) (-1)^{h_S} Q \;\;\; , \;\;\; \e_S(K)(-1)^{h_S} = e^{\pi\i(h_S + h_K -
h_{SK})} =e^{\pi i Q_S(K)} \;\;\; , 
\eeq
where we used that $h_S$ is integer and $\s:=\s_0\s_1$. 
 Thus, the phase  of
the central charge is thus given by minus  
the  automorphism type of the
 crosscap state (since $h_S=1$).    
\end{itemize}
Let us discuss the BPS-relation in some more detail. Explicitly the
mass and the charge equal, up to a common normalization 
\begin{eqnarray}
  \label{eq:3}
  M = \s_0\left(P_{K,O} + \s P_{SK,O}\right)
\;\;\; , \;\;\; Q = \s_0\left(P_{K,S} + \s P_{SK,S}\right) ~.
\end{eqnarray}
Next we note that the identity~\R{PJ} implies that
\beq \label{eq:nul}
P_{NS,R} = 0
\eeq
for any superconformal algebra $\cA$, as can be seen by taking for $J$ the supercurrent $V$.
Because $S\in R$ we therefore have 
\beq
M = \left\{\begin{array}{cc}
    \s_0 P_{OO} & {\rm for} \;\;\; K\in NS \\
    \s_0\s P_{O,KS} & {\rm for} \;\;\; K\in R
    \end{array} \right.~,~~Q = \left\{\begin{array}{cc}
    \s_0\s P_{SK,S} & {\rm for} \;\;\; K\in NS \\
    \s_0 P_{K,S} & {\rm for} \;\;\; K\in R
    \end{array} \right.~.
\eeq
Note that the value of $M$ depends on the spin
structure of $K$. The signs $\s_0,\s$ determine whether the O-planes are O$^+$ or O$^-$
planes.\footnote{We adopt the odd convention that an
 O$^+$ plane has negative tension and a O$^-$ plane has positive
tension.}  For instance, when $K\in NS$ and $P_{00} >0$, $\s_0=-1$ represents O$^+$-planes
and $\s_0=1$ represents O$^-$-planes. Also note that, 
for a fixed choice of $\s_0$ with $K \in NS$, a sign flip in $\s$ 
changes O-planes in anti-O-planes (and vice versa for $K \in R$).

We wish to emphasize 
that it is the proper inclusion of the uncompactified space-time theory 
that guarantees the BPS property. In other words, a single crosscap state
 of the $D_{8-n,1}\times 
\cA^{cy}$ theory of~\cite{Fuchs:2000gv} defined in
the end of section~\ref{sec-CY}, where the GSO projection is only halfway
implemented, is {\em not} BPS.   Let $u\equiv v^ns^2$ again 
denote the simple current that extends
$\cA_{3n}$ to $\cA^{cy}$. A crosscap state of this extension is
\beq \label{eq:Cr1}
|\tilde{\C} \rangle = \sqrt{N_u} \sum_{\{i|REP_{[i]}, Q_u(i)=0\}}
\(\frac{P_{i0} + P_{i,u}}{2\sqrt{S_{0i}}}\)|[i]\rangle\rangle_{\e_u(0)\s} \;\;\; .
\eeq
where $i$ are primaries of $D_{8-n,1}\times 
\cA^{CY}$. (There are more crosscap state due to sign choices and KBC's. The
 point we want to make applies to all of them.) The mass equals $M = \langle 0;\bar{0} | \tilde{\C} \rangle$
and the charge equals $Q = \langle S;\bar{S} | \tilde{\C} \rangle$. Because $u\in NS$ we have $Q=0$ due
to~\R{nul}. Since $M$ generically nonzero, the BPS condition is violated. Another crosscap
state is 
\beq \label{eq:Cr2}
|\tilde{\C} \rangle_S = \sqrt{N_u} \sum_{\{i|REP_{[i]}, Q_u(i)=0\}}
\(\frac{P_{iS} + P_{i,Su}}{2\sqrt{S_{0i}}}\)
 |[i]\rangle\rangle_{\e_u(S)\s} \;\;\; .
\eeq
where we used $S$ as a KBC. Now the mass vanishes and the charge does not. In a sense, the
sum of~\R{Cr1} and~\R{Cr2} obey a BPS condition. (this is the proposal
made in \cite{Govindarajan:2003vv}. A similar proposal for minimal
models was made in \cite{Brunner:2003zm}) 
This sum is precisely what we get
in~\R{Cr}, when we
extend the theory $\cA^{ws}$ by $S$, {\em including} its action 
on the uncompactified space-time part.

With our knowledge of simple current extensions, it is also straightforward to 
see why the theory $D_{8-n,1} \times \cA^{cy}$ does yield BPS D-branes. 
The reason is that the boundary labels, i.e. the modular $S$-matrix, of 
the extended theory $\cA^{ext} \supset D_{8-n,1} \times \cA^{cy}$ is given 
by a representative of the $S$-matrix of $D_{8-n,1} \times \cA^{cy}$. This 
is the content of eq.~\R{BS}. Therefore all properties of the boundary states
are indeed encoded in $\cA^{cy}$. However as the extension is by an even 
current, the $P$-matrix is {\em not} given by a representative, the 
consequences of which we have explicitly shown and discussed above. 

\section{Example: A-type Orientifolds of Gepner Models}
\label{sec:example:-13-gepner}
\setcounter{equation}{0}

When $\cA_{3n}$ is based on $\cN=2$ minimal models the invariant
$\cZ(\cA^{ws},S)$ is called a Gepner model. To illustrate the details
of the computation of BPS CY-crosscaps and how the BPS relation arises, we compute the explicit mass
and charge properties of a crosscap state in a Gepner
model example. Recall that an $\cN=2$ minimal model
$\cA_k$ at level $k$ has a representation as a $SU(2)_k \times
U(1)_4/U(1)_{2h}$ WZW coset; $h\equiv k+2$. It has primaries $(l,m,s)$
with $l=0,\ldots,k$, 
$m=-h+1,\ldots,h$ mod $2h$, $s=-1,\ldots,2$ mod 4, subject to
field identification $(l,m,s) = (k-l,m+h,s+2)$ and selection rule
$l+m+s \in 2 \ZZ$. Field identification in fact 
corresponds to extension by
the identification current $I = (k,h,2)$ on the unconstrained 
primaries $(l,m,s)$ 
\cite{Schellekens:1989uf}. The supercurrent is
$v = (0,0,2)$ and spectral flow is realized by $s = (0,1,1)$. Two
other noteworthy simple currents are the {\em phase symmetries}
generated by $(0,2,0)$ of order $h$,
 and the current $(0,k+2,2)$; the latter 
is the only current
that has fixed points and then only when $k$ is even
\cite{Fuchs:2000fd}. 
We proceed stepwise. First we construct the P-matrix of an
  $\cN=2$ minimal model. We extend the tensor products of minimal
  models to a $\cN=2$ superconformal theory $\cA_{3n}$. Finally we
  apply the lessons from the previous section to compute the PSS O-plane
  mass and charge for odd $k_i$ Gepner models.

\subsection{The $P$-matrix in minimal models}

Using that field identification can be viewed as a simple current
extension, we can express 
the modular $P$ matrix of a coset CFT $G/H$ in terms of
those of the $G$ and $H$ theories with the use of equation~\R{Pext}. There is
one subtlety here. In general we only know the conformal weights in the coset
modulo integers. We can compensate for our ignorance by the introduction of
signs \cite{coset}
\beq
a_{(l,m,s)} := e^{\pi\i [h_{(l,m,s)}^{\it true} - (h_l - h_m + h_s)]}~,
\eeq   
where 
\beq
h_l = \frac{l(l+2)}{4h} \;\;\;,\;\;\; h_m = \frac{m^2}{4h} \;\;\; ,\;\;\;
h_s = \frac{s^2}{8}.
\eeq
For most explicit models, the signs $a$ are known. For minimal model
cosets they can 
be found in for instance~\cite{Brunner:2003zm}.

With these compensatory signs, 
the $P$ matrix of a minimal model at level $k$ is
\beq
P^{min,k}_{(l,m,s)(l',m',s')} = a_{(l,m,s)}a_{(l',m',s')} \sum_{n= 0,1}
\e_{k^n}(l) \e^*_{h^n}(m) \e_{2^n} (s) P^{SU(2)_k}_{l + n(k-2l),l'} \[
P^{U(1)_{2h}}_{m + nh,m'} \]^* P^{U(1)_{4}}_{s + n2,s'} ~;
\eeq
the basic WZW $P$-matrices are
\begin{eqnarray}
  \label{eq:24}
  P^{SU(2)_k}_{l,l'}  &=&
  \ove{\sqrt{h}}\sin\left(\frac{\pi(l+1)(l'+1)}{2h}\right)
  \sum_{u=0}^1 (-1)^{u(k+l+l')} ~,~~~h=k+2 \non
  P^{U(1)_{2p}}_{m,m'} 
&=& \frac{e^{-\frac{\pi i mm'}{2p}}}{2\sqrt{p}}
  \sum_{u=0}^1 (-1)^{u(p+m+m')}~.
\end{eqnarray}

\subsection{The $P$-matrix of $\cA_{3n}$}

Next we construct the tensor product 
\beq 
\cA_{3n}^{ten} := \cA_{k_1} \otimes
... \cA_{k_r} 
\eeq 
with central charge 
$c=\sum_i 3k_i/(k_i+2)= 3n$. We will take all $k_i$ odd to avoid
fixed point ambiguities (though fixed points should have no qualitative effect
on the crosscap states) and label the primaries by
$(\vec{l};\vec{m};\vec{s}):=
((l_1,m_1,s_1),(l_2,m_2,s_2),...,(l_r,m_r,s_r))$.  The $P$-matrix of this
tensor product is simply the product of the minimal model $P$-matrices 
\bea
P^{ten}_{(\vec{l};\vec{m};\vec{s}),(\vec{l}';\vec{m}';\vec{s}')} & = & 
\prod_{i=1}^r \[ P^{min,k_i}_{(l_i,m_i,s_i),(l_i',m_i',s_i')}\]~.
\eea
The currents  $w_i = (v_1,0,...,0,v_i,0,..0)$,
$i=2,...,r$ have integer spin and can be used to extend $\cA_{3n}^{ten}$ to a
$\cN=2$ superconformal algebra $\cA_{3n}$. We denote its primaries by
$[\vec{l};\vec{m};\vec{s}]$. The modular matrices of $\cA_{3n}$
can be expressed in terms of those of the minimal models using the rules~\R{S}
and~\R{PextG} above. In particular, the $P$-matrix of $\cA_{3n}$
  is~\footnote{This expression is correct when the arguments have the lowest
conformal weight modulo $2$ in the orbit. See the remarks around
equation~\R{Pext}.}
\beq 
P^{3n}_{[\vec{l};\vec{m};\vec{s}],[\vec{l}';\vec{m}';\vec{s}']} =
\sum_{\stackrel{\vec{p}}{p_j=0,1;\,\, j=2,...,r}}
\e_{w_2^{p_2}...w_r^{p_r}}((\vec{l};\vec{m};\vec{s}))
P^{ten}_{(w_2^{p_2}...w_r^{p_r}(\vec{l};\vec{m};\vec{s})),(\vec{l}';\vec{m}';\vec{s}')} \;\;\; .
\eeq
For notational purposes it is convenient to define 
\beq
p_1 := \left\{ \begin{array}{ll}
        0 & {\rm for}~\sum_{i=2}^r p_i~ \;{\rm even} \\
        1 & {\rm for}~\sum_{i=2}^r p_i~ \;{\rm odd}
          \end{array}
       \right.
\eeq
Writing out all the phases $a_{(l,m,s)}$ and
$\eps_{\vec{\ome}}(\vec{l},\vec{m},\vec{s})$, the total phase
simplifies and one finds for the $P$-matrix of the $\cA_{3n}$ theory
\bea
\nonumber
 P^{3n}_{[\vec{l};\vec{m};\vec{s}],[\vec{l}';\vec{m}';\vec{s}']} = \hspace{-0.3in}
 \sum_{\mbox{\scriptsize  $
        \begin{array}{c}
        \vec{n},\vec{p}\\
       n_i =0,1;\, i=1,...,r; \\
       p_j=0,1;\, j=2,...,r.\end{array}$}} \hspace{-0.3in}
\[ \prod_{i=1}^r \( a_{(l'_i,m'_i,s'_i)}\)  
\phi_{n_i,p_i}((l_i,m_i,s_i))
P^{SU(2)_{k_i}}_{l_i + n_i(k_i-2l_i),l_i'} 
\[ P^{U(1)_{2h_i}}_{m_i + n_i h_i,m_i'} \]^* 
P^{U(1)_{4}}_{s_i + 2 n_i + 2p_i,s_i'} \]
 ,
\eea
with the phase 
\begin{eqnarray}
  \label{eq:17}
  \phi_{n_i,p_i}((l_i,m_i,s_i)) &\equiv&
  \eps_{\ome_i^{p_i}}((l_i,m_i,s_i))a_{(l_i,m_i,s+2p_i)}\eps_{k_i^{n_i}}(l_i)\eps_{h_i^{n_i}}^{\ast}(m_i)\eps_{2_i^{n_i}}(s_i+2p_i)
  \non
 &=& e^{\pi i \[ h_{(l_i,m_i,s_i)}^{\it true} - h_{l_i+n_i(k-2l_i)}+h_{m_i+2n_ih_i}-h_{s_i+2n_i+2p_i}\]}~.
\end{eqnarray}

\subsection{Masses and charges of O-planes in $\cA^{ws}_{\it Gepner}$}

Using this explicit expression for the $P$-matrix of the $\cA_{3n}$
built from minimal models, we can calculate the mass $M$ and charge $Q$ of the
O-planes for odd 
level Gepner models.
We have explained in detail, that a BPS relation is only obtained
after (i) we tensor the $\cA^{3n}$
theory with 
the space-time part
described by the $D_{8-n,1}$ model, and (ii)
this tensor product is extended by the vector current
$(V_{8-n},v_{3n})$ to
$\cA^{ws}$. 
For concreteness, we consider the PSS-crosscap state, i.e. we choose a trivial Klein bottle current.
To find the charges and tension of the PSS-orientifold plane 
we only need to know
the entries 
$P^{ws}_{OO}$ and $P_{SS}^{ws}$. 
For a Gepner model built on $r$ minimal models with each $k_i$ odd, i.e. models
without fixed points, the computation is straightforward and given in
appendix B. One
finds that these entries of the $P$-matrix are given by 
\begin{eqnarray}
\label{eq:25b}
P^{ws}_{OO} &=& P_{00}^DP^{3n}_{00}+ \eps_V(0) P_{0v}^DP_{0v}^{3n} \non
&=& 2^{r/2}
\cos(\frac{r\pi}{4})(P_{00}^{D}+\tan(\frac{r\pi}{4})
P_{0v}^D) \prod_{i=1}^r\left(P^{min, k_i}_{00}\right)~,
\non
P_{S,S}^{ws} &=& P^D_{s,s}P^{3n}_{s,s}+
  \eps_{V}(s)P^{D}_{s,vs}P^{3n}_{s,v_1s} \non
  &=& 2^{r/2}\cos(\frac{r\pi}{4}) \left(P^D_{s,s} 
  - i\tan(\frac{r\pi}{4}) P^D_{s,vs}\right) \prod_i\left(
  P_{ss}^{min, k_i}\right) ~.
\end{eqnarray}
By~\R{nul}, the mixed $NS$, $R$ entry $P_{O,S}^{ws}$ of vanishes:
$P_{O,S}^{ws}=0$. The non-zero entries of the $P$-matrix of $D_{8-n,1}$ are readily computed (see for instance
\cite{Angelantonj:2002ct})
\begin{eqnarray}
  \label{eq:2}
  P^{D}_{0,0} &=& - P^{D}_{v,v}
  = \cos(\frac{n\pi}{4}) ~,\non
 P^{D}_{0,v} &=&
   -\sin(\frac{n\pi}{4}) ~,\non
 P^{D}_{s,s} &=& P^{D}_{c,c} = e^{\frac{\i n\pi}{4}}\cos(\frac{n\pi}{4}) ~,\non
 P^{D}_{s,c} &=& -\i e^{\frac{\i n\pi}{4}}\sin(\frac{n\pi}{4})~.
\end{eqnarray}
Substituting these values, we get
\begin{eqnarray}
  \label{eq:4}
  P^{ws}_{O,O} &=& 2^{r/2}
\cos(\frac{(n+r)\pi}{4})
\prod_{i=1}^r\left(P^{min, k_i}_{00}\right) ~,\non
  P^{ws}_{S,S} &=& 2^{r/2}
\cos(\frac{(n+r)\pi}{4})e^{\frac{\i n \pi}{4}}
\prod_{i=1}^r\left(P^{min, k_i}_{ss}\right) = 2^{r/2}
\cos(\frac{(n+r)\pi}{4})
\prod_{i=1}^r\left(P^{min, k_i}_{00}\right) ~,
\end{eqnarray}
upon using (\ref{eq:27}) and $n =\sum k_i/(k_i+2)$.

As $P_{S,S}^{ws}$ {\em equals} $P_{O,O}^{ws}$ we immediately see that the PSS-crosscap state has equal magnitude 
mass and central
  charge: 
\begin{eqnarray}
  \label{eq:6}
  M = \langle O | C\rangle^+_{O} = \sig_0 P_{O,O}
  &,& 
  Q = \langle S | C\rangle^+_{O} = \sig_0 P_{S,S}~,
  \non 
  M = \langle O | C\rangle^-_{O}  = \sig_0 P_{O,O}&,& 
  Q = \langle S | C\rangle^-_{O}  = - \sig_0 P_{S,S}~.
\end{eqnarray}
The BPS-like equality is verified, and the different choices for the
automorphism sign $\sig$  
are explicitly seen to correspond to planes
vs. anti-planes. The result, of course, is an explicit manifestation of
the identity~\R{PJ} when we recall that $h_S=1$.

\section{Conclusions}

The results of~\cite{FOE}
 have opened the way for a systematic study of a large class of
open unoriented rational CFTs.
Of special
phenomenological interest are those unoriented 
RCFTs which correspond to type I string
compactifications on 'rational' Calabi-Yau manifolds. 
These theories provide a
new class of $\cN=1$, $d=4$ string vacua, in addition 
to heterotic strings on
Calabi-Yau 3-folds and M-theory on $G_2$ manifolds. In RCFT-based
 string theories, spacetime
 supersymmetry is implemented through a GSO
 projection. This projection can equivalently 
be viewed as a simple current extension by the
 spectral flow current $S$. 
Using the general theory of boundary and crosscap states for
simple current extensions~\cite{FOE,thesis}, we have
constructed here the crosscap states for such unoriented 
type I compactifications.
These boundary and crosscap states are expressed in terms of the
\mbox{(pseudo-)}~modular
matrices $S$ and $P$, which are explicitly
known for many CFTs, e.g. 
WZW models and cosets thereof.\footnote{Since the conformal weights
  $h$ of coset theories are in general only known up to integers, the
  coset $P$-matrix is in general only 
 known up to a sign.}

In particular, we have shown
how the rational CY crosscap states~\R{Cross} correspond to 
half-supersymmetry preserving A-type orientifold planes that are
BPS-like. Their masses and central charges are equal up to a phase: 
$M=e^{i\phi}Q$, and this relation is a reflection of the simple current
identity (\ref{eq:PJ}) obeyed by the modular $P$-matrix. 
The phase $\phi$, moreover, is minus the automorphism-type
of the crosscap state with respect to the spectral flow current.
This BPS condition only holds
when the uncompactified spacetime
degrees of freedom are properly included in the GSO projection.
This is in contrast with D-branes~\cite{walcher}, where the BPS
property follows from considering the internal sector independent of
the space-time sector.

This study provides a step towards the classification of 
the orientifolds
of a given Calabi-Yau manifold. RCFT methods are
limited to CY manifolds at rational points in the moduli space, and
ultimately one wishes for a geometric
description where one can freely move away from 'rational'
Calabi-Yaus. Progress towards a geometric formulation of orientifolds
is in  
the early stages
\cite{Acharya:2002ag,Blumenhagen:2002wn,Misra:2003zv,Diaconescu:2003dq}. A recent study of unoriented linear sigma models
\cite{Brunner:2003zm} showed that orientifold planes 
are located at fixed
points of holomorphic or anti-holomorphic isometries. A next item is
to determine their charges and tension. Matching with RCFT data, as
obtained with the methods described here, can provide these.

\bigskip
\noindent
{\bf Acknowledgments:} We thank Brandon Bates, Charles Doran, Suresh
Govindarajan, Brian
Greene, Jaydeep Majumder, 
and Bert Schel\-le\-kens for help and useful comments.
A number of calculations were checked with Bert Schellekens' program
{\tt kac} (http://www.nikhef.nl/$\sim$t58/kac.html), an indispensable
tool.
KS is grateful for partial support from DOE
grant DE-FG-02-92ER40699. LH wants to thank the theory groups of both the
NIKHEF 
in Amsterdam, The Netherlands, and Columbia University, New York 
for their hospitality.

\appendix

\section{Crosscap states in simple current extensions}
\setcounter{equation}{0}
\label{sec:autom-type-simple}

To show explicitly some of the properties of the 
crosscap state of a simple current extension, we perform here 
the steps discussed in section \ref{sec:bound-crossc-stat} in detail.
Starting with the natural guess eq.~(\ref{eq:CROSS}) for a $\ZZ_{2m}$ 
simple
current crosscap state (for ease of notation we have absorbed the normalization $1/\sqrt{2m
  S_{0i}}$ in the crosscap Ishibashi state)
\beq 
\label{eq:CROSSb}
|\C \rangle^{[\s]}_{[K]} = \sum_{n=0}^{2m-1} \sum_i 
\sig(J^nK) P_{i,J^nK} |i\rangle\rangle_{1,C}~,
\eeq
we use the simple current identity~\R{PJ} for even
and odd powers $J^n$ respectively to obtain
\begin{eqnarray}
  \label{eq:7}
  |\C \rangle^{[\s]}_{[K]} &=& \sum_i \sum_{n=0}^{m-1} \left( \sig(J^{2n}K) P_{i,J^{2n}K} +
   \sig(J^{2n+1}K) P_{i,J^{2n+1}K} \right)|i\rangle\rangle_{1,C}  \non
&=& \sum_i \sum_{n=0}^{m-1} \left( \sig(J^{2n}K)
   \eps_{J^{2n}}(K)e^{2\pi\i[Q_{J^n}(i)-Q_{J^n}(J^nK)]}P_{i,K} +
   \right. \non
 && ~~~~~~~~\left. + \sig(J^{2n+1}K) \eps_{J^{2n}}(JK)e^{2\pi\i[Q_{J^n}(i)-Q_{J^n}(J^{n+1}K)]}
   P_{i,JK} \right)|i\rangle\rangle_{1,C} ~.
\end{eqnarray}
Making the sign choices advocated in eq~\R{signs}, 
\begin{eqnarray}
  \label{eq:8}
\s(J^{2m}K) \e_{J^{2m}}(K) e^{-2\pi\i Q_{J^m}(J^mK)} & = & \s_0 \;\;\;,
\non
\s(J^{2m+1}K) \e_{J^{2m}}(JK) e^{-2\pi\i Q_{J^m}(J^{m+1}K)} & = & \s_1
\;\;\;  ,
\end{eqnarray}
we get
\begin{eqnarray}
  \label{eq:9}
  |\C \rangle^{[\s]}_{[K]} &=& \sum_i (\sig_0 P_{i,K}+\sig_1 P_{i,JK})
   \sum_{n=0}^{m-1} e^{2\pi\i Q_{J^n}(i)} |i\rangle\rangle_{1,C}~.
\end{eqnarray}
The final sum over phases $e^{2\pi\i Q_{J^n}(i)}$ 
immediately shows that fields with monodromy charge
$Q_J(i)= 2k/2m,~k\neq 0$ do not couple to the orientifold plane. Nor do odd
charged fields, for in that case both $P$-matrix entries vanish. For a
$\ZZ_{2m}$ current the $P$-matrix obeys
\begin{eqnarray}
 P_{a,i} &=& P_{J^{2m}a,i} \non
&=& \eps_{J^{2m}}(a)e^{2\pi\i[Q_J^m(i)-Q_J^m(J^ma)]}P_{a,i} \non
&=& e^{2\pi\i m [Q_{J}(i)-Q_J(a)]} P_{a,i}~.
\end{eqnarray}
We used in 
the second line again the identity~\R{PJ}; in
the third line the definitions for $\eps_J(K)$ and $Q_J(K)$, the fact
that $J$ has integer weight, and that $Q_{J^p}(J^qi)=p(Q_J(i)+qQ_J(J))$ mod 1.
From this identity it follows that a $P$-matrix element with an even and and
odd charged entry vanishes: for $Q_{J}(i) = 2k+1/2m$ and
$Q_J(a)=2p/2m$,  $P_{a,i}=-P_{a,i}=0$. As the current $K$ must be even
in order that eq. (\ref{eq:8}) can be solved (see eq. \ref{eq:even}),
hence for odd charged fields both $P$-matrix entries $P_{i,K}=0$ and
$P_{i,JK}$ vanish.  
This establishes the claim that the sign choice
made, ensures that the O-plane only couples to fields in the
extension, i.e. fields with $Q_J(i)= 0$ mod 1. 

We may therefore limit
the sum over primaries to the sum over chargeless primaries with no
penalty.
The final step is to rewrite the sum over $\cA$ primaries, as a sum
over representatives of chargeless orbits $[i]_{Q_J(i)=0}$, the
primaries of the extended theory:
\begin{eqnarray}
  \label{eq:10}
  |\C \rangle^{[\s]}_{[K]} &=& \sum_{\left\{i|Q_J(i)=0\right\}} (\sig_0
   P_{i,K}+\sig_1 P_{i,JK}) |i\rangle\rangle_{1,C} \non
&=& \sum_{\{i|REP_{[i]},~Q_J(i)=0\}}\sum_{n=0}^{2m-1} (\sig_0
   P_{J^ni,K}+\sig_1 P_{J^ni,JK}) |J^ni\rangle\rangle_{1,C}~.
\end{eqnarray}
Splitting the sum into even and odd parts, and using the $P$-matrix
simple current identity once more, we find
\begin{eqnarray}
  \label{eq:11}
  |\C \rangle^{[\s]}_{[K]}  =
\sum_{\{i|REP_{[i]},~Q_J(i)=0\}}\sum_{n=0}^{m-1}&& \left[
~~~\sig_0 \,\eps_{J^{2n}}(i) ~e^{2\pi\i[Q_{J^n}(K)-Q_{J^n}(J^ni)]}~~P_{i,K}
  \right.|J^{2n}i\rangle\rangle_{1,C} \\[-.1in] \nonumber
&& ~+\sig_0\,
  \eps_{J^{2n}}(Ji)e^{2\pi\i[Q_{J^n}(K)-Q_{J^n}(J^{n+1}i)]}P_{Ji,K}
|J^{2n+1}i\rangle\rangle_{1,C} 
 \non
&& ~+\sig_1 \,
  \eps_{J^{2n}}(i) ~~e^{2\pi\i[Q_{J^n}(JK)-Q_{J^n}(J^ni)]}P_{i,JK}
|J^{2n}i\rangle\rangle_{1,C}
 \non
&& \left. ~+ \sig_1\,
  \eps_{J^{2n}}(Ji)e^{2\pi\i[Q_{J^n}(JK)-Q_{J^n}(J^{n+1}i)]}P_{Ji,JK}
 |J^{2n+1}i\rangle\rangle_{1,C} \right].
\end{eqnarray}
An alternative form of the $P$-matrix simple current identity~\R{PJ} is~\cite{thesis}
\begin{eqnarray}
  \label{eq:12}
  P_{Jd,b} = \eps^{\ast}_J(d)\eps_J(b)P_{d,Jb}~.
\end{eqnarray}
With this we see that 
\begin{eqnarray}
  \label{eq:13}
  P_{Ji,K} &=& \eps^{\ast}_J(i)\eps_J(K) P_{i,JK}~, \non
  P_{Ji,JK} &=& \eps^{\ast}_J(i) \eps_J(JK)P_{i,J^2K} = \eps^{\ast}_J(i)
  \eps_J(JK) \eps_{J^2}(K) e^{2\pi\i[Q_J(i)-Q_J(JK)]}P_{i,K} \non
&=&
 \eps^{\ast}_J(i) \eps_J(K)e^{2\pi\i[Q_J(i)-h_J]} P_{i,K}~,
\end{eqnarray}
where we used~\R{PJ} and the definitions of $\eps_J(K)$ and $Q_J(K)$ 
in the second and third step.
Using that the field $i$ is chargeless: $Q_J(i)$ is zero, and that $J$
is a simple current used in an extension, i.e. it has integer
conformal weight, one can show that $\eps^{\ast}_J(i)
=\eps_J(i)$. Substituting these identities above we find
\begin{eqnarray}
  \label{eq:15}
    |\C \rangle^{[\s]}_{[K]}  =
\sum_{\{i|REP_{[i]},~Q_J(i)=0\}}\sum_{n=0}^{m-1}&& \left[
~~(\sig_0 P_{i,K}+\sig_1 P_{i,JK}) ~\eps_{J^{2n}}(i)e^{2\pi\i Q_{J^n}(K)}~~|J^{2n}i\rangle\rangle_{1,C}
  \right. \\[-.1in] \nonumber
&& \left. +(\sig_0\,P_{i,JK} +\sig_1P_{i,K})
  \eps_{J^{2n}}(Ji)\eps_J(i)\eps_J(K) e^{2\pi\i Q_{J^p}(K)}P_{i,JK}
|J^{2n+1}i\rangle\rangle_{1,C} \right].
\end{eqnarray}
where we have again used that $Q_{J^p}(J^qi)= p(Q_J(i)+qQ_J(J))$ mod 1, and
that $Q_J(J)=0$ mod 1. Finally realizing that for an integer spin
current $J$, $e^{2\pi i Q_J(K)}=\eps_J^2(K)$, and recombining of the
explicit expressions for the various $\eps_J(i)$ terms, we can simplify the
expression for the crosscap state to
\begin{eqnarray}
  \label{eq:16}
  |\C \rangle^{[\s]}_{[K]}  &=&
\sum_{\{i|REP_{[i]},~Q_J(i)=0\}} (\sig_0 P_{i,K}+\sig_1P_{i,JK})
\sum_{n=0}^{m-1} \left[ \eps_J^{2n}(K)\eps_{J^{2n}}(i) |J^{2n}i\rangle\rangle_{1,C}
 \right. \non
&& \hspace{2.7in} \left.+ \sig \eps_J^{2n+1}(K) \eps_{J^{2n+1}}(i)|J^{2n+1}i\rangle\rangle_{1,C} \right] \non
&=& \sum_{\{i|REP_{[i]},~Q_J(i)=0\}} (\sig_0 P_{i,K}+\sig_1P_{i,JK}) \sum_{n=0}^{2m-1}
\left[ \sig \eps_J(K)\right]^{n}\eps_{J^{n}}(i) |J^{n}i\rangle\rangle_{1,C} ~.
\end{eqnarray}
This is eq.~\R{Cross}.

\section{$P$-matrix entries in all $k_i$ odd Gepner models} 
\setcounter{equation}{0}
\label{sec:p-matrix-entries}

\begin{itemize}
\item $P_{OO}^{ws}$

Recall the forms of the $SU(2)$ and $U(1)$ WZW $P$-matrices,
eq.~(\ref{eq:24}). 
Due to the selection rule $k+l_1+l_2 \in 2\ZZ$ for the $SU(2)$
$P$-matrix, the $P_{00}$ and $P_{0v}$ entries of the $P$-matrix of the 
odd $k$ minimal models are given by a single term 
(recall the conformal weights
$h_{k,h,2}=0$,~$h_{0,0,2}=h_{k,k+2,0}=3/2$, see e.g. 
\cite{Brunner:2003zm}) 
\begin{eqnarray}
  \label{eq:1}
  P^{min}_{00}&=& \frac{\sqrt{2}}{(k+2)}
  \sin\left(\frac{\pi(k+1)}{2k+4}\right)~, \non
P^{min}_{0v} &=& P^{min}_{00}~.
\end{eqnarray}
Tensoring $r$ odd $k_i$ minimal models 
and extending by $w_i$ 
we obtain for the $P_{00}$ and $P_{0v}$ elements 
of the $\cA^{3n}$ theory (use that
$\eps_v(0)=e^{-3\pi\i/2}$)\footnote{Compared to the expression above
  eq. (\ref{eq:17}), which computes $P^{3n}_{00}$ in one step, we have
  first performed the sum over $n$ to obtain the minimal model
  $P$-matrix, and then extended by the currents $w_i$. In this second
  step, we used that
  $$ \eps_{w_2^{p_2}\ldots w_r^{p_r}}(\vec{l};\vec{m};\vec{s})  =
  \prod_{i=1}^r\eps_{w_i^{p_i}}((l_i,m_i,s_i))~, $$ which is only true
  if one has chosen the correct representative of the orbit. 
  In general one has to be careful with the
  additional phases $a_{(l_i,m_i,s_i)}$. }
\begin{eqnarray}
  \label{eq:5}
  P^{3n}_{00} &=& \prod_{i=1}^r \left(P_{00}^{min,k_i}\right)+
    \sum_{i<j}\eps_{v_i}(0)P_{0v_i}\eps_{v_j}(0)P_{0v_j}\prod_{k\neq
    1,j}\left(P_{00}^{k_k}\right)+ \non
&&~~~~~~+
    \sum_{i<j<k<l}
    \eps_{v_i}(0)P_{0v_i}\eps_{v_j}(0)P_{0v_j}\eps_{v_k}(0)P_{0v_k}\eps_{v_l}(0)P_{0v_l}\prod_{n
    \neq i,j,k,l} \left(P_{00}^{k_n}\right)+\ldots
    \non 
&=&
\sum_{p=0}^{r/2}(-1)^p\left(\matrix{ r\cr 2p}
    \right)
   \prod_{i=1}^r \left(P_{00}^{min,k_i}\right)\non
&=& 
    2^{r/2}\cos(\frac{r\pi}{4}) \prod_{i=1}^r
    \left(P_{00}^{min,k_i}\right) ~,\non
  P^{3n}_{0v} &=& \sum_{p=0}^{r/2} (-1)^p\left(\matrix{ r \cr
    2p+1}\right)
    \prod_i\left(P_{00}^{min,k_i}\right) \non
&=& 2^{r/2}\sin(\frac{r\pi}{4}) \prod_i\left(P_{00}^{min,k_i}\right)~.
\end{eqnarray}
Finally extending with $V$ we get (note that $h_V=2$, hence $\eps_V(O)=1$)
\begin{eqnarray}
\label{eq:25}
P^{ws}_{OO} &=& P_{00}^DP^{3n}_{00}+ \eps_V(0) P_{0v}^DP_{0v}^{3n} \non
&=& 2^{r/2}
\cos(\frac{r\pi}{4})(P_{00}^{D}+\tan(\frac{r\pi}{4})
P_{0v}^D) \prod_{i=1}^r\left(P^{min,k_i}_{00}\right)~.
\end{eqnarray}

\item $P_{O,S}^{ws}$

From the explicit expression 
\begin{eqnarray}
\label{eq:it2}
P^{min}_{(\lam,\mu,\sig),(l,m,s)} =
P^{SU(2)}_{\lam,l}\left(P^{U(1)}_{\mu,m}\right)^{\ast}
P^{U(1)}_{\sig,s} + e^{\pi\i[h_{l,m,s}-h_{k-l,m+h,s+2}]}P^{SU(2)}_{\lam,k-l}\left(P^{U(1)}_{\mu,m+h}\right)^{\ast}
P^{U(1)}_{\sig,s+2}~,
\end{eqnarray}
it is easy to see that $P^{min}_{0,s}= 0$ for $k$ odd. The selection
rule that $k+a+b \in 2\ZZ$ for $SU(2)_k$ $P$-matrices
implies that in that case only the second term contributes, but the
same selection rule $k+a+b \in 2\ZZ$ for the $U(1)_{2k+_4}$ factor
then shows that $P^{min}_{00}=0$. Since the currents $v_i$ only act in
the $U(1)_4$ sector, this immediately shows $P^{ws}_{O,S}=0$. Of
course, this is simply an example of $P_{NS,R}=0$

\item $P_{S,S}^{ws}$

For the entry $P^{min}_{ss}$ again only the second term in
(\ref{eq:it2}) contributes due to the $SU(2)$ selection rule
\begin{eqnarray}
  \label{eq:27}
  P^{min}_{(0,1,1),(0,1,1)} = e^{\pi\i[h_{(0,1,1)}-h_{(k,-k-1,-1)}]}
P^{SU(2)}_{0,k}\left(P^{U(1)}_{1,-k-1}\right)^{\ast}
P^{U(1)}_{1,-1} 
&=& e^{-\frac{\pi\i k}{4k+8}}P_{00}^{min} ~.
\end{eqnarray}
We also need
\begin{eqnarray}
\label{eq:14}
P^{min}_{s,vs} = -e^{-\frac{\pi i(3k+4)}{4k+8}}P_{00}^{min}  = i
P_{s,s}^{min}~.
\end{eqnarray}
Extending with the currents $w_i$ we find ($\eps_v(s)=-1$)
\begin{eqnarray}
  \label{eq:19}
  P^{3n}_{ss} &=& \prod_{i=1}^r\left(P_{ss}^{min,k_i}\right)+
    \sum_{i<j}\eps_{v_i}(s)P_{s,v_is}\eps_{v_j}(s)P_{s,v_js}\prod_{k\neq
    i,j}\left(P_{ss}^{k_k}\right) +\non
&&~~~~~+
    \sum_{i<j<k<l}\eps_{v_i}(s)P_{s,v_is}\eps_{v_j}(s)P_{s,v_js}\eps_{v_k}(s)P_{s,v_ks}\eps_{v_l}(s)P_{s,v_ls}\prod_{n\neq i,j,k,l}
    \left(P_{ss}^{k_n}\right)+\ldots
    \non 
&=&
\sum_{p=0}^{r/2}(-1)^p\left(\matrix{ r\cr 2p}
    \right)
   \prod_i\left(P_{ss}^{min,k_i}\right) \non
&=& 2^{r/2}\cos(\frac{r\pi}{4}) \prod_i\left(P_{ss}^{min,k_i}\right)~,
\non
P^{3n}_{s,v_1s} &=& i 2^{r/2}\sin(\frac{r\pi}{4})\prod_i\left(P_{ss}^{min,k_i}\right)~.
\end{eqnarray}
Extending finally with $V$, we get, using $\eps_{V}(S)= -1$, 
\begin{eqnarray}
  \label{eq:29}
  P_{S,S}^{ws} &=& P^D_{s,s}P^{3n}_{s,s}+
  \eps_{V}(s)P^{D}_{s,vs}P^{3n}_{s,v_1s} \non
  &=& 2^{r/2}\cos(\frac{r\pi}{4}) \left(P^D_{s,s} 
  - i\tan(\frac{r\pi}{4}) P^D_{s,vs}\right) \prod_i\left(
  P_{ss}^{min,k_i}\right)~.
\end{eqnarray}
\end{itemize}

\end{document}